\newcounter{myctr}
\def\myitem{\refstepcounter{myctr}\bibfont\noindent\ifnum\themyctr>9\else\phantom{0}\fi\hangindent17pt\themyctr.\enskip}

\documentclass{ws-ijqi}

\usepackage{braket}
\usepackage{kbordermatrix}

\newtheorem{thm}{Theorem}

\begin{document}

\bibliographystyle{unsrt}

\markboth{Takuya Machida}
{Phase transition of a continuous-time quantum walk on the half line}

\title{Phase transition of a continuous-time quantum walk on the half line}

\author{Takuya Machida}

\address{College of Industrial Technology, Nihon University, Narashino, Chiba 275-8576, Japan\\
machida.takuya@nihon-u.ac.jp}

\maketitle

\begin{abstract}
Quantum walks are referred to as quantum analogs to random walks in mathematics.
They have been studied as quantum algorithms in quantum information for quantum computers.
There are two types of quantum walks.
One is the discrete-time quantum walk and the other is the continuous-time quantum walk.
We study a continuous-time quantum walk on the half line and challenge to find a limit theorem for it in this paper.
As a result, approximate behavior of the quantum walker is revealed after the system of quantum walk gets updated in long time.
\end{abstract}

\keywords{Quantum walk, Limit theorem, Localization}

\section{Introduction}
Related to quantum algorithms, quantum walks have been investigated as quantum analogs to random walks in mathematics~\cite{Venegas-Andraca2012}.
Since a continuous-time quantum walk was introduced in 1998~\cite{FarhiGutmann1998} for quantum computation on trees, continuous-time quantum walks have been studied analytically and numerically.
Given Hamiltonians, quantum walkers spread on discrete spaces (i.e. graphs) under time evolution rules which are described by Schr\"{o}dinger equations.
One of the goals in mathematical study for quantum walks is to discover where the walkers locate in space after the systems of quantum walks get updated in long time.  
Long-time limit theorems have been focused on so that we understand the asymptotic behavior of the walkers after their long time evolutions.
While we see a lot of limit theorems for discrete-time quantum walks, few limit theorems, which describe the asymptotic behavior of continuous-time quantum walks in specific, have been proved.
Using one of the results which Avraham {\it et al.}~\cite{Avraham2004} found for some continuous-time quantum walks by Laplace transform, Konno derived a long-time limit distribution of a continuous-time quantum walk on the line in 2005~\cite{Konno2005b}.
He also proved a limit distribution for a continuous-time quantum walk on trees in 2006~\cite{Konno2006}.
Machida~\cite{Machida2023} studied a continuous-time quantum walk on the line whose Hamiltonian was spatially 2-periodic, and got a limit distribution in 2023.
Only convergence in distribution was proved for more general models in some papers~\cite{Gottlieb2005,BoutetdeMonvelSabri2023}.

We are going to analyze a continuous-time quantum walk on the half line $\mathbb{Z}_{\geq 0}=\left\{0,1,2,\ldots\right\}$ and try to provide a long-time limit theorem for it in this paper.
The continuous-time quantum walk is defined by spatially 2-periodic Hamiltonian, which is similar to the Hamiltonian given for the quantum walk on the line.
A limit distribution for the quantum walk on the line was proved by Fourier analysis in Machida~\cite{Machida2023}.
But, it is difficult to analyze the quantum walk on the half line in a similar method because the representation of the Schr\"{o}dinger equation at the edge is different from the ones at the other locations.
Therefore, we will try to get a limit theorem by another method, that is Laplace transform.

We define a continuous-time quantum walk on the half line with a Schr\"{o}dinger equation in Sect.~\ref{sec:definition}.
The Hamiltonian is determined by two real numbers, represented by $\gamma_0$ and $\gamma_1$, and the initial state of the walker is supposed to localize at the edge of the half line.
The main result is provided in Sect.~\ref{sec:limiting_amplitude}, which is concluded as a limit theorem for the finding probability of the walker.
Then, we will discuss about phase transition of the continuous-time quantum walk between localization and delocalization in Sect.~\ref{sec:summary}.
After the main contents, we will see numerical experiments for a continuous-time quantum walk on the finite line in~\ref{app:numerics}, so that we compare our main result to the numerical experiments.
\ref{app:invariant_state} is devoted for the invariant state of the Schr\"{o}dinger equation, which is briefly mentioned in Sect.~\ref{sec:summary}.

\section{A continuous-time quantum walk on the half line}
\label{sec:definition}
The system of continuous-time quantum walk at time $t\,(\,\geq 0)$ is described by probability amplitude $\left\{\psi_t(x)\in\mathbb{C} : x\in\mathbb{Z}_{\geq 0}\right\}$, where $\mathbb{C}$ is the set of complex numbers.
We assume that the quantum walker launches off at the edge of the half line,
\begin{equation}
 \psi_0(x)=\left\{\begin{array}{ll}
  1 & (x=0)\\
	    0 & (x=1,2,\ldots)
	   \end{array}\right..\label{eq:initial}
\end{equation}

\begin{figure}[h]
 \begin{center}
  \includegraphics[scale=0.6]{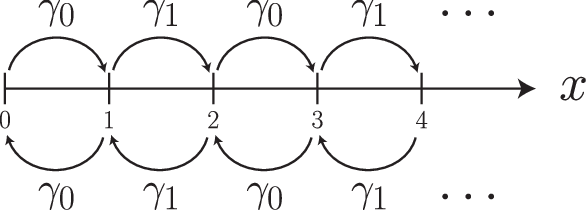}
  \caption{The Hamiltonian of the Schr\"{o}dinger equations Eqs.~\eqref{eq:time_ev_0}, \eqref{eq:time_ev_even}, and \eqref{eq:time_ev_odd}.}
  \label{fig:1}
 \end{center}
\end{figure}

Let $\gamma_0$ and $\gamma_1$ be real numbers.
The probability amplitude at time $t$ gets updated in a Schr\"{o}dinger equation
\begin{align}
 i\,\frac{d}{dt}\psi_t(0)=&\, \gamma_0\,\psi_t(1),\label{eq:time_ev_0}\\
 i\,\frac{d}{dt}\psi_t(2n)=&\, \gamma_1\,\psi_t(2n-1)+\gamma_0\,\psi_t(2n+1)\qquad (n=1,2,\ldots),\label{eq:time_ev_even}\\
 i\,\frac{d}{dt}\psi_t(2n+1)=&\, \gamma_0\,\psi_t(2n)+\gamma_1\,\psi_t(2n+2)\qquad (n=0,1,2,\ldots),\label{eq:time_ev_odd}
\end{align}
where $i$ denotes the imaginary unit.
With the matrix form of the Schr\"{o}dinger equation
\begin{equation}
i\,\frac{d}{dt}
\kbordermatrix{
& \\
& \psi_t(0)\\
& \psi_t(1)\\
& \psi_t(2)\\
& \psi_t(3)\\
& \psi_t(4)\\
& \vdots
}
=
\kbordermatrix{
 & 0 & 1 & 2 & 3 & 4 & \cdots \\
0 &  0 &  \gamma_0 & 0 &  0 &  0  &  \cdots \\
1 &  \gamma_0 &  0 & \gamma_1 &  0 &  0  & \cdots \\
2 &  0 &  \gamma_1 & 0 & \gamma_0 & 0 & \cdots \\
3 &  0 &  0 & \gamma_0 & 0 & \gamma_1 & \cdots \\
4 &  0 &  0 & 0 & \gamma_1 & 0 & \cdots \\
\vdots & \vdots & \vdots  & \vdots & \vdots & \vdots & \ddots 
}
\kbordermatrix{
& \\
& \psi_t(0)\\
& \psi_t(1)\\
& \psi_t(2)\\
& \psi_t(3)\\
& \psi_t(4)\\
& \vdots
},
\end{equation}
we clearly see the Hamiltonian
\begin{equation}
\kbordermatrix{
 & 0 & 1 & 2 & 3 & 4 & \cdots \\
0 &  0 &  \gamma_0 & 0 &  0 &  0  &  \cdots \\
1 &  \gamma_0 &  0 & \gamma_1 &  0 &  0  & \cdots \\
2 &  0 &  \gamma_1 & 0 & \gamma_0 & 0 & \cdots \\
3 &  0 &  0 & \gamma_0 & 0 & \gamma_1 & \cdots \\
4 &  0 &  0 & 0 & \gamma_1 & 0 & \cdots \\
\vdots & \vdots & \vdots  & \vdots & \vdots & \vdots & \ddots 
}.
\end{equation}
Figure.~\ref{fig:1} would be helpful to understand the Hamiltonian.

\noindent
We assume that the values of parameters $\gamma_0$ and $\gamma_1$ are not zero (i.e. $\gamma_0, \gamma_1\neq 0$).

Let $X_t$ be the position of the quantum walker at time $t$.
The walker is observed at position $x$ at time $t$ with probability
\begin{equation}
 \mathbb{P}(X_t=x)=\bigl|\psi_t(x)\bigr|^2.
\end{equation}

\section{Limiting amplitude as $t\to\infty$}
\label{sec:limiting_amplitude}

Let us define the Laplace transform of the amplitude,
\begin{equation}
 F_x(s)=\int_{0}^{\infty}e^{-st}\psi_t(x)\,dt.
\end{equation}
The differential equations~\eqref{eq:time_ev_0}, \eqref{eq:time_ev_even}, and \eqref{eq:time_ev_odd} are transformed into recurrence relations with regard to only the positions,
\begin{align}
 i\,\Bigl(\,sF_{0}(s)-\psi_0(0)\,\Bigr)=&\, \gamma_0\,F_{1}(s),\label{eq:L_ev_0_1}\\
 i\,\Bigl(\,sF_{2n}(s)-\psi_0(2n)\,\Bigr)=&\, \gamma_1\,F_{2n-1}(s)+\gamma_0\,F_{2n+1}(s)\qquad (n=1,2,\ldots),\label{eq:L_ev_even_1}\\
 i\,\Bigl(\,sF_{2n+1}(s)-\psi_0(2n+1)\,\Bigr)=&\, \gamma_0\,F_{2n}(s)+\gamma_1\,F_{2n+2}(s)\qquad (n=0,1,2,\ldots).\label{eq:L_ev_odd_1}
\end{align}
Remembering the initial state in Eq.~\eqref{eq:initial}, we organize the recurrence relations,
\begin{align}
 F_{1}(s)=&\, i\,\frac{\,sF_{0}(s)-1\,}{\,\gamma_0\,},\label{eq:L_ev_0_2}\\
 F_{2n+1}(s)=&\,\frac{\,is\,}{\,\gamma_0\,}F_{2n}(s)-\frac{\,\gamma_1\,}{\,\gamma_0\,}F_{2n-1}(s)\qquad (n=1,2,\ldots),\label{eq:L_ev_even_2}\\
 F_{2n+2}(s)=&\,\frac{\,is\,}{\,\gamma_1\,}F_{2n+1}(s)-\frac{\,\gamma_0\,}{\,\gamma_1\,}F_{2n}(s)\qquad (n=0,1,2,\ldots).\label{eq:L_ev_odd_2}
\end{align}
With the same representation as Eq.~\eqref{eq:L_ev_odd_2},
\begin{equation}
  F_{2n}(s)=\,\frac{\,is\,}{\,\gamma_1\,}F_{2n-1}(s)-\frac{\,\gamma_0\,}{\,\gamma_1\,}F_{2n-2}(s)\qquad (n=1,2,\ldots),\label{eq:L_ev_odd_3}
\end{equation}
we make vector forms from Eqs.~\eqref{eq:L_ev_even_2} and \eqref{eq:L_ev_odd_3},
\begin{align}
 \begin{bmatrix}
  F_{2n+1}(s)\\ F_{2n}(s)
 \end{bmatrix}
 =&
 \begin{bmatrix}
  \frac{\,is\,}{\,\gamma_0\,} & -\,\frac{\,\gamma_1\,}{\,\gamma_0\,}\\
  1 & 0
 \end{bmatrix}
 \begin{bmatrix}
  F_{2n}(s)\\ F_{2n-1}(s)
 \end{bmatrix}\qquad (n=1,2,\ldots),\\
 \begin{bmatrix}
  F_{2n}(s)\\ F_{2n-1}(s)
 \end{bmatrix}
 =&
 \begin{bmatrix}
  \frac{\,is\,}{\,\gamma_1\,} & -\,\frac{\,\gamma_0\,}{\,\gamma_1\,}\\
  1 & 0
 \end{bmatrix}
 \begin{bmatrix}
  F_{2n-1}(s)\\ F_{2n-2}(s)
 \end{bmatrix}\qquad (n=1,2,\ldots),
\end{align}
from which
\begin{equation}
 \begin{bmatrix}
  F_{2n+1}(s)\\ F_{2n}(s)
 \end{bmatrix}
 =
 (M_1M_0)^n
 \begin{bmatrix}
  F_{1}(s)\\ F_{0}(s)
 \end{bmatrix}\qquad (n=1,2,\ldots),
\end{equation}
follows, where
\begin{equation}
 M_0=
 \begin{bmatrix}
  \frac{\,is\,}{\,\gamma_1\,} & -\,\frac{\,\gamma_0\,}{\,\gamma_1\,}\\
  1 & 0
 \end{bmatrix},\quad
 M_1=
 \begin{bmatrix}
  \frac{\,is\,}{\,\gamma_0\,} & -\,\frac{\,\gamma_1\,}{\,\gamma_0\,}\\
  1 & 0
 \end{bmatrix}.
\end{equation}
Using the eigenvalues of the matrix $M_1M_0$, represented by $q_{\pm}(s)$, we get
\begin{align}
 & (M_1M_0)^n\nonumber\\
 =&\,\frac{\,1\,}{\,2\,}\biggl\{\,q_{+}(s)^n+q_{-}(s)^n+\frac{\,s^2-\gamma_0^2+\gamma_1^2\,}{\,\sqrt{\,p(s)\,}\,}\Bigl(\,q_{+}(s)^n-q_{-}(s)^n\,\Bigr)\,\biggr\}
 \begin{bmatrix}
  1 & 0\\ 0 & 0
 \end{bmatrix}\nonumber\\
 &\,+i\,\frac{\,s\gamma_0\,}{\,\sqrt{\,p(s)\,}\,}\Bigl(\,q_{+}(s)^n-q_{-}(s)^n\,\Bigr)
 \begin{bmatrix}
  0 & 1\\ 0 & 0
 \end{bmatrix}\nonumber\\
 &\,-i\,\frac{\,s\gamma_0\,}{\,\sqrt{\,p(s)\,}\,}\Bigl(\,q_{+}(s)^n-q_{-}(s)^n\,\Bigr)
 \begin{bmatrix}
  0 & 0\\ 1 & 0
 \end{bmatrix}\nonumber\\
 &+\,\frac{\,1\,}{\,2\,}\biggl\{\,q_{+}(s)^n+q_{-}(s)^n-\frac{\,s^2-\gamma_0^2+\gamma_1^2\,}{\,\sqrt{\,p(s)\,}\,}\Bigl(\,q_{+}(s)^n-q_{-}(s)^n\,\Bigr)\,\biggr\}
 \begin{bmatrix}
  0 & 0\\ 0 & 1
 \end{bmatrix},
\end{align}
where
\begin{align}
 q_{\pm}(s)=&\,-\frac{\,\Bigl(\,\sqrt{\,s^2+(\gamma_0+\gamma_1)^2\,}\,\pm\,\sqrt{\,s^2+(\gamma_0-\gamma_1)^2\,}\,\Bigr)^2\,}{\,4\gamma_0\gamma_1\,},\\
 p(s)=&\,\Bigl\{s^2+(\gamma_0+\gamma_1)^2\Bigr\}\Bigl\{s^2+(\gamma_0-\gamma_1)^2\Bigr\}.
\end{align}
Note that $q_{+}(s)q_{-}(s)=1$.
Looking back at Eq.~\eqref{eq:L_ev_0_2}, we find, for $n=1,2,\ldots$,
\begin{align}
 F_{2n+1}(s)=&\,\frac{\,i\,}{\,\sqrt{\,p(s)\,}\,}\biggl\{\Bigl(\gamma_0+\gamma_1 q_{+}(s)-s\gamma_1 F_{0}(s)q_{+}(s)\Bigr)q_{+}(s)^n\nonumber\\
&\qquad\qquad -\Bigl(\gamma_0+\gamma_1 q_{-}(s)-s\gamma_1 F_{0}(s)q_{-}(s)\Bigr)q_{-}(s)^n\biggr\},\\
 F_{2n}(s)=&\,\frac{\,1\,}{\,\sqrt{\,p(s)\,}\,}\biggl[-\Bigl\{s+\gamma_1 F_{0}(s)\Bigl(\gamma_1+\gamma_0 q_{+}(s)\Bigr)\Bigr\}q_{+}(s)^n\nonumber\\
 &\qquad\qquad +\Bigl\{s+\gamma_1 F_{0}(s)\Bigl(\gamma_1+\gamma_0 q_{-}(s)\Bigr)\Bigr\}q_{-}(s)^n\biggr].
\end{align}

Here, let the value of $s$ fix at a positive real number.
Then we are going to find the initial state of the Laplace transform. 
For $s>0$, since $|q_{+}(s)|$ is larger than 1 and $F_{x}(s)$ is bounded,
\begin{equation}
 \big|F_{x}(s)\bigr|\,=\,\left|\,\int_{0}^{\infty}e^{-st}\psi_t(x)\,dt\,\right|
 \,\leq \,\int_{0}^{\infty}e^{-st}\,\bigl|\psi_t(x)\bigr|\,dt
 \,\leq \,\int_{0}^{\infty}e^{-st}\,dt
 \,=\,\frac{\,1\,}{\,s\,},
\end{equation}
the coefficient of $q_{+}(s)^n$ should be $0$, from which $F_{0}(s)$ follows,
\begin{equation}
 F_{0}(s)=\,\frac{\,\gamma_1+\gamma_0 q_{-}(s)\,}{\,s\gamma_1\,}\,=\,-\frac{\,s q_{-}(s)\,}{\,\gamma_1(\gamma_0+\gamma_1 q_{-}(s))\,}.
\end{equation}
As a result, we get the precise representation of the Laplace transform,
\begin{align}
 F_{2n}(s)=&\,\frac{\,\gamma_1+\gamma_0 q_{-}(s)\,}{\,s\gamma_1\,}\,q_{-}(s)^n \quad (n=1,2,\ldots),\\
 F_{2n+1}(s)=&\,\frac{\,i\,}{\,\gamma_1\,}\,q_{-}(s)^{n+1}\quad (n=1,2,\ldots),
\end{align}
which are correct for $n=0$.
We should note that $|q_{-}(s)|<1\,\,(s>0)$ because of $q_{+}(s)q_{-}(s)=1$.
The final value theorem tells us the long-time limit of the amplitude.
For $n=0,1,2\ldots$, we get
\begin{align}
 \lim_{t\to\infty}\psi_t(2n)=& \lim_{s\to +0}sF_{2n}(s)\nonumber\\
 =&\left\{\begin{array}{cl}
    {\displaystyle \left\{\,1-\left(\frac{\,\gamma_0\,}{\,\gamma_1\,}\right)^2\,\right\} \left(-\frac{\,\gamma_0\,}{\,\gamma_1\,}\right)^n} & (\,|\gamma_0| < |\gamma_1|\,)\\[5mm]
     0 & (\,|\gamma_0| \geq |\gamma_1|\,)
	  \end{array}\right.,\\[3mm]
 \lim_{t\to\infty}\psi_t(2n+1)=& \lim_{s\to +0}sF_{2n+1}(s)\,=\,0,
\end{align}
from which a limit theorem for the probability distribution follows.\\
\begin{thm}
 The quantum walker launches off at the localized initial state given in Eq.~\eqref{eq:initial}.
 For $n=0,1,2,\ldots$, we have
 \begin{align}
  \lim_{t\to\infty}\mathbb{P}(X_t=2n) & = \left\{\begin{array}{cl}
					   {\displaystyle \left\{\,1-\left(\frac{\,\gamma_0\,}{\,\gamma_1\,}\right)^2\,\right\}^2 \left(\frac{\,\gamma_0\,}{\,\gamma_1\,}\right)^{2n}} & (\,|\gamma_0| < |\gamma_1|\,)\\[5mm]
					    0 & (\,|\gamma_0| \geq |\gamma_1|\,)
						 \end{array}\right.,\\[3mm]
  \lim_{t\to\infty}\mathbb{P}(X_t=2n+1) & = 0.
 \end{align}
 \label{th:limit}
\end{thm}

Since the total sum of the limit,
\begin{equation}
 \sum_{x=0}^\infty\, \lim_{t\to\infty}\mathbb{P}(X_t=x)
  =\left\{\begin{array}{cl}
    {\displaystyle 1-\left(\frac{\,\gamma_0\,}{\,\gamma_1\,}\right)^2} & (\,|\gamma_0| < |\gamma_1|\,)\\[5mm]
     0 & (\,|\gamma_0| \geq |\gamma_1|\,)
	  \end{array}\right.,
\end{equation}
is less than $1$, $\lim_{t\to\infty}\mathbb{P}(X_t=x)$ is not a probability distribution.

\section{Summary}
\label{sec:summary}

We studied a continuous-time quantum walk on the half line $\mathbb{Z}_{\geq 0}$ and found the long-time limit of the amplitude via Laplace transform.
The continuous-time quantum walk was operated by a discrete-space Schr\"{o}dinger equation with a spatially 2-periodic Hamiltonian which was determined by two real numbers $\gamma_0$ and $\gamma_1$.
The amplitude vanished all over the positions as $t\to\infty$, if $|\gamma_0|\geq |\gamma_1|$.
On the other hand, it vanished at positions $x=1,3,5,\ldots$, and rested at positions $x=0,2,4,\ldots$, as $t\to\infty$, if $|\gamma_0|<|\gamma_1|$.

\begin{figure}[h]
 \begin{center}
  \includegraphics[scale=0.6]{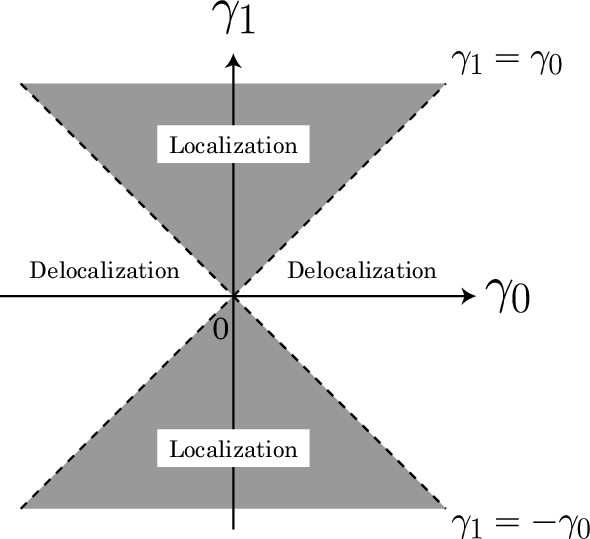}
  \caption{Phase transition between localization and delocalization. Dot lines are included in delocalization area.}
  \label{fig:2}
 \end{center}
\end{figure}

Let us define the localization condition for the continuous-time quantum walk as $\exists\,x\in\mathbb{Z}_{\geq 0},\, \lim_{t\to\infty}\mathbb{P}(X_t=x)\,>\,0$.
Then, our result is concluded in a phase transition of the continuous-time quantum walk between localization and delocalization, as shown in Fig.~\ref{fig:2}.

The limit of the amplitude is an invariant state of the Schr\"{o}dinger equation (See \ref{app:invariant_state}), but it is not the invariant amplitude because $\sum_{x=0}^\infty \bigl|\,\lim_{t\to\infty}\psi_t(x)\,\bigr|^2\,\neq\, 1$.
The limit, therefore, gives the limiting measure, not the limiting distribution.
Limiting distributions for discrete-time or continuous-time quantum walks have been proved, most of which are described as convergence in law for walker's position rescaled by time $t$ as $t\to\infty$, and they are useful to understand the approximated behavior of the quantum walkers as time $t$ is large enough.
The limiting measure is a first step to get an approximation, but we still have a challenging work on the limiting distribution so that we understand the continuous-time quantum walk more precisely.

\bigskip

\begin{center}
{\bf Acknowledgements}
\end{center}
The author is supported by JSPS Grant-in-Aid for Scientific Research (C) (No. 23K03220).


\appendix
\section{Continuous-time quantum walk on the finite line}
\label{app:numerics}

Let $N$ be a positive even integer, which will be fixed at $N=500$ through this section.
We define a continuous-time quantum walk on the finite line $\mathbb{Z}_{N}=\left\{0,1,2,\ldots, N-1\right\}$.
The amplitude $\psi_t(x)\,(x\in\mathbb{Z}_N)$ gets updated by a Schr\"{o}dinger equation,
\begin{align}
 i\,\frac{d}{dt}\psi_t(0)=& \gamma_0\,\psi_t(1),\\
 i\,\frac{d}{dt}\psi_t(2n)=& \gamma_1\,\psi_t(2n-1)+\gamma_0\,\psi_t(2n+1)\nonumber\\
 & \qquad\qquad (n=1,2,\ldots, N/2-1),\\[1mm]
 i\,\frac{d}{dt}\psi_t(2n+1)=& \gamma_0\,\psi_t(2n)+\gamma_1\,\psi_t(2n+2)\nonumber\\
 & \qquad\qquad (n=0,1,2,\ldots, N/2-2),\\[1mm]
 i\,\frac{d}{dt}\psi_t(N-1)=& \gamma_0\,\psi_t(N-2),
\end{align}
starting with the localized initial state at one of the edges,
\begin{equation}
 \psi_0(x)=\left\{\begin{array}{ll}
  1 & (x=0)\\
	    0 & (x=1,2,\ldots, N-1)
	   \end{array}\right..
\end{equation}

We discretize the Schr\"{o}dinger equation regarding to time $t$ and use an alternate model to experiment the continuous-time quantum walk in numeric, 
\begin{align}
 i\,\frac{\,\phi_{t+\Delta t}(0)-\phi_{t}(0)\,}{\Delta t} =& \gamma_0\,\phi_t(1),\\
 i\,\frac{\,\phi_{t+\Delta t}(2n)-\phi_{t}(2n)\,}{\Delta t} =& \gamma_1\,\phi_t(2n-1)+\gamma_0\,\phi_t(2n+1)\nonumber\\
 & \qquad\qquad (n=1,2,\ldots, N/2-1),\\[1mm]
 i\,\frac{\,\phi_{t+\Delta t}(2n+1)-\phi_{t}(2n+1)\,}{\Delta t} =& \gamma_0\,\phi_t(2n)+\gamma_1\,\phi_t(2n+2)\nonumber\\
 & \qquad\qquad (n=0,1,2,\ldots, N/2-2),\\[1mm]
 i\,\frac{\,\phi_{t+\Delta t}(N-1)-\phi_{t}(N-1)\,}{\Delta t} =& \gamma_0\,\phi_t(N-2),
\end{align}
with the initial state
\begin{equation}
 \phi_0(x)=\left\{\begin{array}{ll}
  1 & (x=0)\\
	    0 & (x=1,2,\ldots, N-1)
	   \end{array}\right..
\end{equation}
Aiming at comparing $|\phi_t(x)|^2$ to the limit which we found for the continuous-time quantum walk on the half line in Sect.~\ref{sec:limiting_amplitude}, we numerically study the alternate model on the finite line $\mathbb{Z}_{500}$ as $\Delta t=0.0001$. 

\begin{figure}[h]
\begin{center}
 \begin{minipage}{50mm}
  \begin{center}
   \includegraphics[scale=0.4]{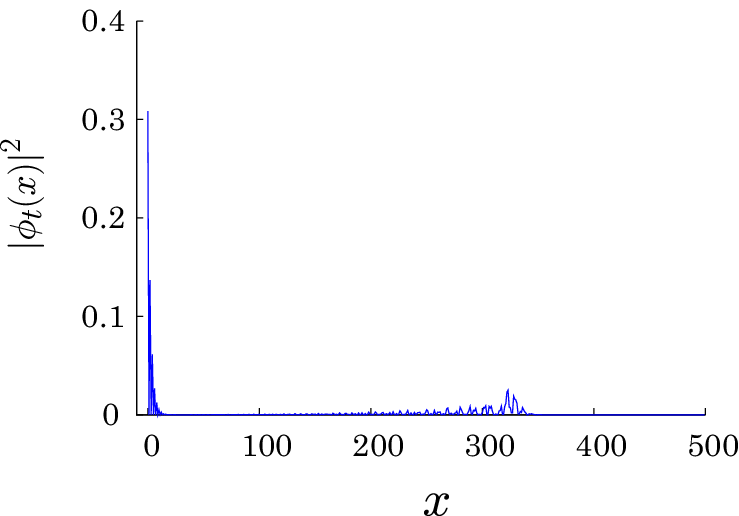}\\[2mm]
  (a) $\gamma_0=1/3,\, \gamma_1=1/2$
  \end{center}
 \end{minipage}\hspace{10mm}
 \begin{minipage}{50mm}
  \begin{center}
   \includegraphics[scale=0.4]{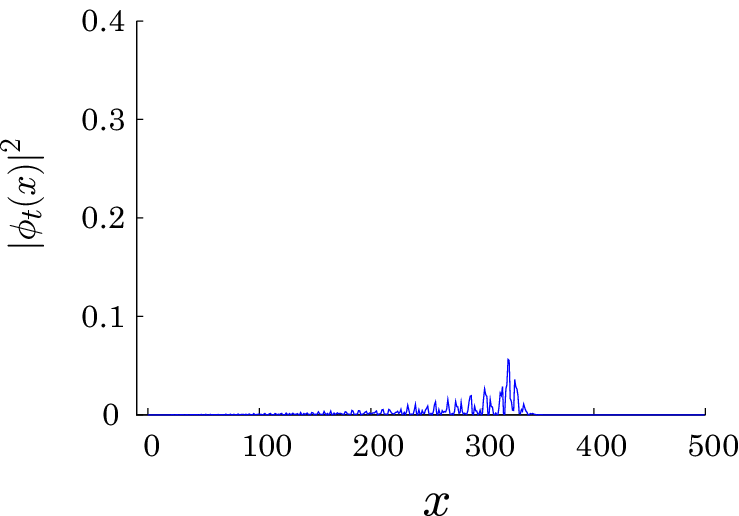}\\[2mm]
  (b) $\gamma_0=1/2,\, \gamma_1=1/3$
  \end{center}
 \end{minipage}
\caption{(Color figure online) The blue lines represent $|\phi_t(x)|^2$ at time $t=500$.}
\label{fig:3}
\end{center}
\end{figure}

\begin{figure}[h]
\begin{center}
 \begin{minipage}{60mm}
  \begin{center}
   \includegraphics[scale=0.3]{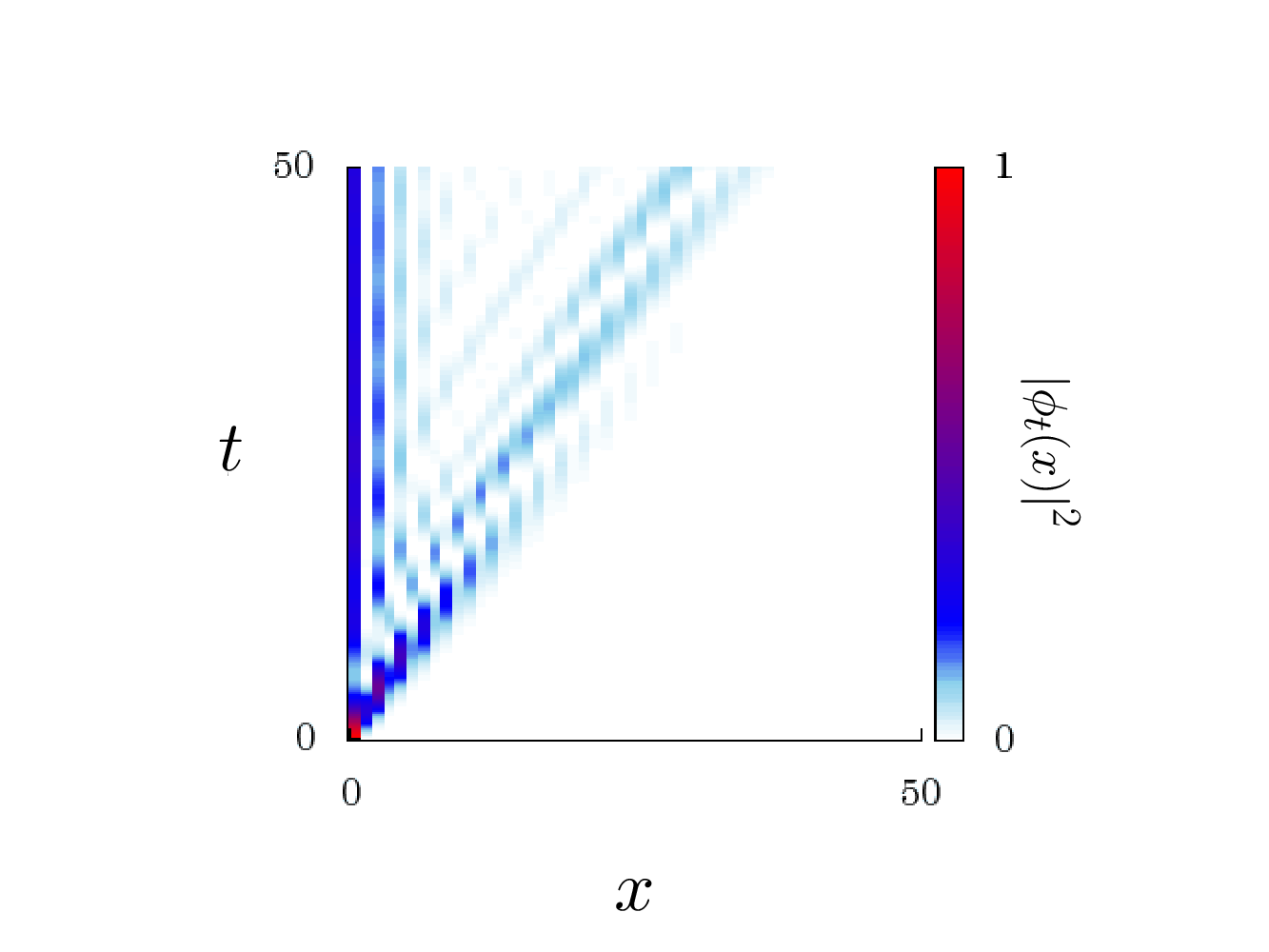}\\[2mm]
  (a) $\gamma_0=1/3,\,\gamma_1=1/2$
  \end{center}
 \end{minipage}
 \begin{minipage}{60mm}
  \begin{center}
   \includegraphics[scale=0.3]{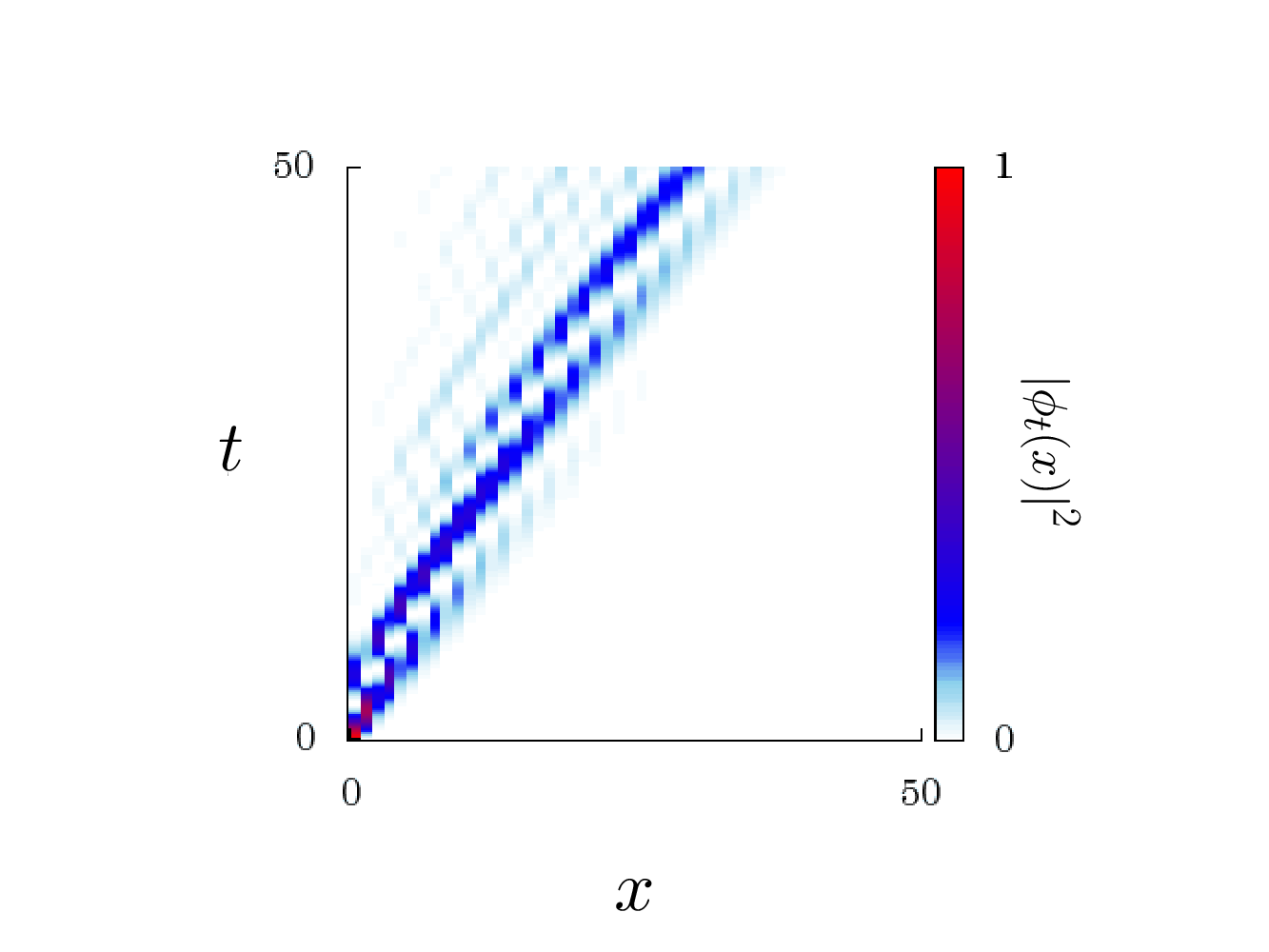}\\[2mm]
  (b) $\gamma_0=1/2,\,\gamma_1=1/3$
  \end{center}
 \end{minipage}
\caption{(Color figure online) $|\phi_t(x)|^2$ shows ballistic behavior.}
\label{fig:4}
\end{center}
\end{figure}

\begin{figure}[h]
 \begin{center}
  \begin{minipage}{60mm}
   \begin{center}
    \includegraphics[scale=0.45]{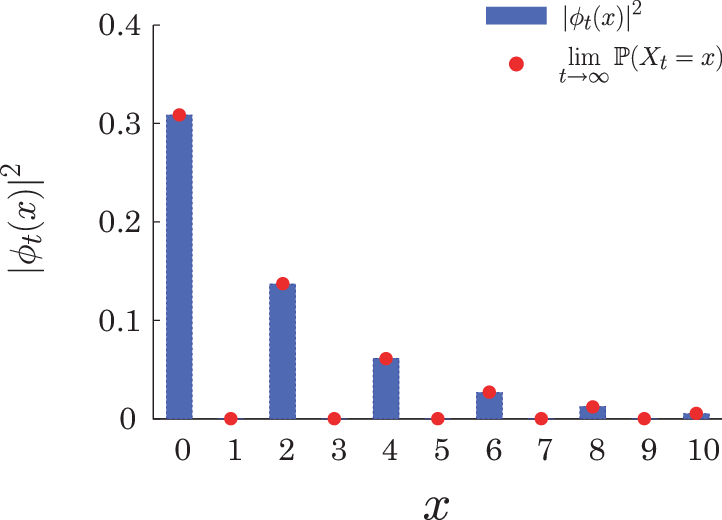}\\
    (a) $\gamma_0=1/3,\,\gamma_1=1/2$
   \end{center}
  \end{minipage}
  \begin{minipage}{60mm}
   \begin{center}
    \includegraphics[scale=0.45]{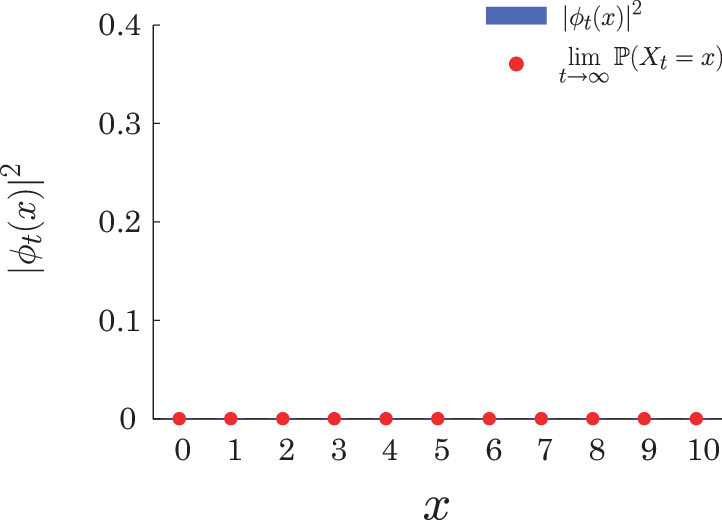}\\
    (b) $\gamma_0=1/2,\,\gamma_1=1/3$
   \end{center}
  \end{minipage}
  \caption{(Color figure online) Comparison between $|\phi_t(x)|^2$ at time $t=500$ and $\lim_{t\to\infty}\mathbb{P}(X_t=x)$.}
  \label{fig:5}
 \end{center}
\end{figure}

\begin{figure}[h]
 \begin{center}
  \begin{minipage}{60mm}
   \begin{center}
    \includegraphics[scale=0.45]{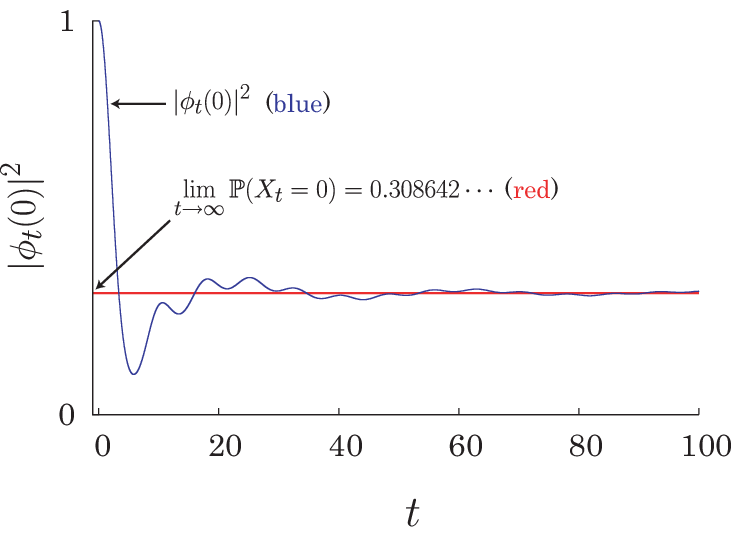}\\
    (a) $\gamma_0=1/3,\,\gamma_1=1/2$
   \end{center}
  \end{minipage}
  \begin{minipage}{60mm}
   \begin{center}
    \includegraphics[scale=0.45]{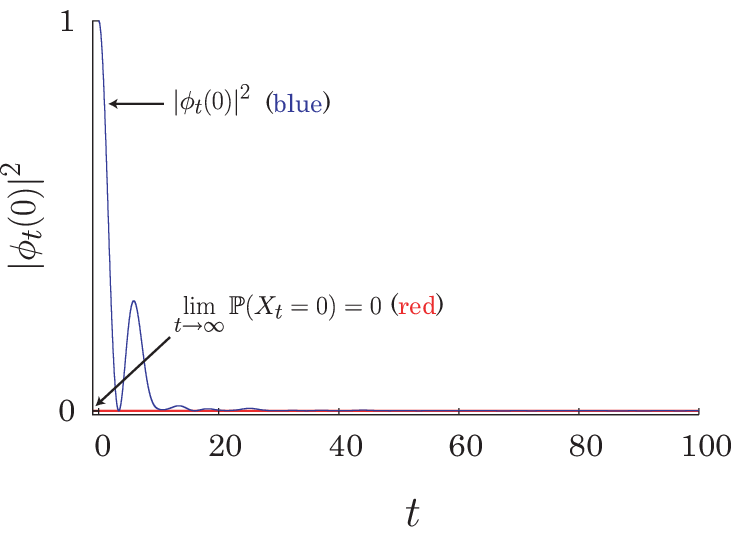}\\
    (b) $\gamma_0=1/2,\,\gamma_1=1/3$
   \end{center}
  \end{minipage}
  \caption{(Color figure online) Comparison between $|\phi_t(0)|^2$ and $\lim_{t\to\infty}\mathbb{P}(X_t=0)$.}
  \label{fig:6}
 \end{center}
\end{figure}

The measure $|\phi_t(x)|^2$ concentrates around the launching point $x=0$ in Fig.~\ref{fig:3}--(a), but such a concentration is not observed in Fig.~\ref{fig:3}--(b).
Figure~\ref{fig:4} explains how the measure $|\phi_t(x)|^2$ spreads, and it shows ballistic behavior.
The measure keeps the concentration around the origin only in Fig.~\ref{fig:4}--(a).
Figures~\ref{fig:5} and \ref{fig:6} compare the numerical experiments to the limit which we found in Sect.~\ref{sec:limiting_amplitude}.
Focusing on the position $x=0$, we see the measure $|\phi_t(0)|^2$ approaching the limiting measure as time $t$ is going up, shown in Fig.~\ref{fig:6}. 

\section{Invariant state}
\label{app:invariant_state}

We find the invariant state $\varphi(x)\,(x=0,1,2,\ldots)$ (\,i.e. $d\varphi(x)/dt=0$\,) of the Schr\"{o}dinger equation in Eqs.~\eqref{eq:time_ev_0}, \eqref{eq:time_ev_even}, and~\eqref{eq:time_ev_odd}.
Solving
\begin{align}
 i\cdot 0=&\, \gamma_0\,\varphi(1),\\
 i\cdot 0=&\, \gamma_1\,\varphi(2n-1)+\gamma_0\,\varphi_t(2n+1)\qquad (n=1,2,\ldots),\\
 i\cdot 0=&\, \gamma_0\,\varphi(2n)+\gamma_1\,\varphi(2n+2)\qquad (n=0,1,2,\ldots),
\end{align}
we have
\begin{equation}
 \varphi(2n)=\,\varphi(0)\left(-\frac{\,\gamma_0\,}{\,\gamma_1\,}\right)^n,\quad
 \varphi(2n+1)=\,0\qquad (n=0,1,2,\ldots).
\end{equation}
There, hence,  exists an invariant state if $\varphi(0)\in \mathbb{C}\,\backslash\left\{\,0\,\right\}$.

The total sum of $\bigl|\varphi(x)\bigr|^2$ exists if $|\gamma_0|<|\gamma_1|$,
\begin{equation}
 \sum_{x=0}^{\infty}\,\bigl|\varphi(x)\bigr|^2
 \,=\,\frac{\,\bigl|\varphi(0)\bigr|^2\,}{\displaystyle \,1-\left(\frac{\,\gamma_0\,}{\,\gamma_1\,}\right)^2\,}.
\end{equation}
The invariant amplitude can be realized by the complex number $\varphi(0)$ such that
\begin{equation}
 \bigl|\varphi(0)\bigr|\,=\,\sqrt{\,1-\left(\frac{\,\gamma_0\,}{\,\gamma_1\,}\right)^2\,},
\end{equation}
under the condition $|\gamma_0|<|\gamma_1|$.

\end{document}